\documentclass[aps,epsfig,twocolumn,pre,showpacs]{revtex4}
\usepackage{graphicx}
\usepackage{amsmath}

\newcommand{\la}{\overleftarrow}
\newcommand{\ra}{\overrightarrow}
\newcommand{\ta}{\overleftrightarrow} 
\newcommand{\be}{\begin{equation}}
\newcommand{\ee}{\end{equation}}
\newcommand{\mn}{\mathcal{N}}
\newcommand{\ml}{\mathcal{L}}

\begin{document}

\title{Inversion method for content-based networks}

\author{Jos\'e J. Ramasco} \email{jramasco@isi.it}
\affiliation{CNLL, ISI Foundation, Viale S. Severo 65, I-10133 Torino, Italy}

\author{Muhittin Mungan} \email{mmungan@boun.edu.tr}
\affiliation{Department of Physics, Faculty of Arts and Sciences,\\
Bo\u gazi\c ci University, 34342 Bebek, Istanbul, Turkey,}
\affiliation{The Feza G\"ursey Institute,
P.O.B. 6, \c Cengelk\"oy, 34680 Istanbul, Turkey,}

\date{\today}

\begin{abstract}

In this paper, we generalize a recently introduced Expectation Maximization 
(EM) method for graphs and apply it
to content-based networks. The EM method provides a classification of the 
nodes of a graph, and allows to infer relations between the different 
classes. Content-based networks are ideal models for graphs displaying any kind of community or/and multipartite structure. We show both numerically and analytically that the generalized EM method is able to recover the process that led to the generation of such networks. 
We also investigate the conditions under which our generalized EM method can recover the underlying 
contents-based structure in the presence of randomness in the connections. Two 
entropies, $S_q$ and $S_c$, are defined to measure 
the quality of the node classification and to what extent the connectivity of a given network is 
content-based. $S_q$ and $S_c$ 
are also useful in determining the number of classes for which the 
classification is optimal.  
\end{abstract}

\pacs{89.75.Hc, 02.50.Tt} 

\maketitle

\section{Introduction}

Classifying items with respect to their properties is a fundamental and very old problem. If the properties are inherent to the objects, the difficulty is deciding first how many groups are required and then establishing the discrimination thresholds for each. The matter becomes more complicated when instead of the inherent properties, one tries to classify elements based on mutual interactions. Of course, such classifications would be very useful for a better understanding of the mechanisms underlying the behavior of systems encountered in scientific disciplines as diverse as Sociology, Biology or Physics \cite{doreian05,freeman04,ravasz02,newman03}. 
As an example, consider social systems which are often modeled as networks. The vertices represent individuals and the edges interactions between them. These interactions can be of many types: friendship, belonging to the same club or school, working together, etc. In these graphs, it is important to be able to group the nodes into what is commonly known as {\it communities}. That is, groups of vertices that share a higher number of connections among themselves than with the rest of the network 
\cite{newman02,newman04,newman06,radicchi,santo07r} (see also \cite{santo07} for a recent review). This partition bears information on which persons have a stronger interdependence and may allow to predict the actors that drive the dynamics of the group as a whole. In Biology, on the other hand, network methods have been used to understand gene regulatory patterns \cite{lee02}. Here each vertex corresponds to a gene and an edge contains information on how the associated protein regulates the synthesis 
of the protein associated to the second gene. Since regulation of gene activity plays a fundamental role in the functioning of the cell \cite{cell}, the community structure points towards the different functional subunits (see \cite{babu04} and references therein). Given the relevance of communities, the last years have seen an increase in the number of techniques proposed to detect them. To name a few: some of them are based on the concept of betweenness (number of paths passing through a link) and modularity \cite{newman-girvan,newman04,newman06}, others on synchronization of oscillators \cite{alex-arenas,boccaletti} or on other dynamical 
systems running on the network \cite{bornholdt,guimera,kumpula}, detection of overlapping 
cliques \cite{palla} or the diffusion of random walkers \cite{zhou,simonsen,gfeller}.

Nevertheless, communities are not the only relevant information that can be extracted from networks. It is also possible to search for vertices with similar connection patterns (not necessarily having connections among themselves, as in the case of communities) 
that are expected to play equivalent functional roles. 
In the social networks literature such nodes are referred to as {\it structurally equivalent} 
\cite{LorrainWhite71} and have lead to an analysis of social networks based on {\it Block Modeling} \cite{whiteboormanbreiger76,doreian05}. In many types of networks, like those formed by webpages or social actors, the connection between nodes is often due to some intrinsic properties of the nodes, which we will refer to henceforth as their "contents". 
Thus it is possible to consider an alternative point of view in which a network structure arises as a result of node contents, leading 
one to the notion of contents-based networks \cite{balcanerzan,mungan,balcan,balcan2}. 

In many cases, network analysis approaches based on communities and those based on some 
form of node similarity are aimed towards the understanding of very different questions. When viewed within 
the framework of contents based networks, however, these differences disappear as will be argued below. We will also show that  
an extension of Newman and Leicht's Expectation Maximization (EM) method  \cite{newman07} is well-suited for uncovering content-based structure underlying a network, inverting in practice the process that lead to its formation.

The organization of the paper is as follows: in Section II, content based networks are formally introduced. Next, we describe in Section III our generalization of the EM method to directed graphs. In Section IV, we show how the EM method can be used to solve 
the inverse problem, namely to recover the underlying contents-based structure from a given network. We present in Section V analytical results regarding the 
application of the EM method to contents-based networks and the recovery of the 
contents-based structure. These results will be complemented with a numerical study in Sections VI and VII. In Section VII,  we consider 
a more realistic situation and ask to what extent an underlying contents-based structure can be recoverred in the presence of disorder in the connections. Finally,  we summarize our results and present the conclusions in Section VIII.

\section{Content Based Networks}

Let us define first content-based networks. Consider a set of nodes 
$i = 1, 2, \ldots N$ each of which has a {\it content} $x_i$ assigned with $x_i \in \mathcal{X} = \{1,2, \ldots, \mathcal{N}_x\}$, and where $1,2, \ldots $ are 
labels for the possible contents. The structure of the connectivity pattern of the associated content-based 
network is determined by the function  $c(x_i,x_j) \in \{0,1\}$, which 
is defined for all ordered pairs of contents $(x,y) \in \mathcal{X}$. 
The adjacency matrix of the graph is then given by 
\begin{equation}
A_{ij} = c(x_i,x_j).
\end{equation}

We see immediately that nodes having the same contents $x$ also have the same connection patterns, and thus are structurally equivalent \cite{LorrainWhite71}. As explained before, this can imply a functional equivalence in the process that generated the network. The 
point of view that we will take in this article is to regard contents-based
networks as ideal networks, from which the "real" networks are obtained through alteration 
or removal of some of the connections. Note that the range of topologies that can be generated via contents-based network is very broad: if the connectivity function $c(x,y)$ shows a close to diagonal configuration, the network will be formed by a set of almost insulated cliques. The ideal configuration would be a family of independent communities without interconnections. Another configuration that can be easily reproduced with content-based networks are bipartite graphs. In its most simplest from, it is enough to allow the nodes to take one of two possible contents and letting the connectivity function 
$c$ to be  non-zero only for the off-diagonal elements. Much more complicated connectivity patterns can be actually achieved by introducing finer contents distinctions and more intricate connectivity functions. Thus a content based graph can in general include all sorts of combinations between community-like and/or multipartite graphs, as can be seen in the example plotted in Figure 1. 

Another point to note is that originally these networks were proposed in a context where the relation between contents was an order relation \cite{farmeretal,weissbuchetal,balcanerzan}. 
This implies that the relation between nodes is not symmetric and the network is therefore more naturally represented by a directed graph. In this case, the connectivity function 
$c$ is non-symmetric in its arguments. Apart from directionality, realistic graphs may present, as well, a certain degree of disorder in their connection patterns. This effect can be incorporated into the mathematical description by regarding the values of $c(x,y)$ as probabilities of having a link from a node of content $x$ to a node of content $y$. This view transforms the content-based network into a hidden variable graph \cite{caldarelli,soderberg,marian}. As we will see later, the EM method is still able to extract information from networks produced in this way but the failure rate increases the further $c(x,y)$ deviates from a binary-valued function.

\begin{figure}
\includegraphics[width=7cm]{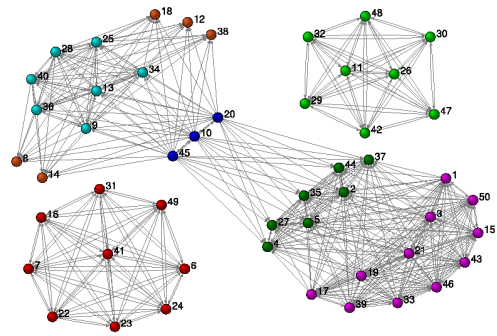}
\caption{An example of content based network, the colors correspond to the different contents (green $A$, red $B$, blue $C$, magenta $D$, cyan $E$, olive $F$ and orange $G$).}
\end{figure}

Contents based networks have proven to be very useful in the description of phenomena that include an underlying relation of hierarchy or ordering. The simplest way of achieving 
such a relation is to associate with each node a string of letters and letting the 
relation between any two nodes be based on string inclusion: namely that one string is 
contained as an uninterrupted subsequence in the other. Networks generated from 
random strings in this manner have been successfully used to model receptor-ligand 
interactions in the immune system \cite{farmeretal,weissbuchetal},  and the transcription factor based gene regulatory network in yeast \cite{balcanerzan, mungan, balcan, balcan2}.

In this article, our goal is to address the {\it inverse} problem: Given a network of 
which we know nothing in advance,  is it possible to decide whether there is an underlying contents-based structure and, if so, can we deduce the class membership of its nodes and the class connectivity function? Moreover, can this be achieved in the presence of noisy connections? Seen in this way, the problem at hand becomes one of statistical inference, very well-suited to EM methods \cite{embook,garlaschelli}.

\section{The EM Method for Networks and its Generalization}
\label{EMGen}

Given a graph $\mathcal{G}$ of $N$ nodes and an adjacency matrix $A_{ij}$, 
the Expectation Maximization (EM) algorithm \cite{newman07} searches for a 
partition of the nodes into $\mathcal{N}_c $ groups such that a certain 
log-likelihood function for the graph is maximized. Henceforth we will refer 
to the groups into which the EM method divides the nodes, as {\it classes}. 
Note that $\mathcal{N}_c $ must not be confused with the number of contents 
$\mathcal{N}_x$,  described in the previous section. Ideally, the optimal number of classes would be $\mathcal{N}_x$, but a criterion independent from the EM algorithm is required to determine its value since we assume that in general 
$\mathcal{N}_x$ will not be known in advance. The variables of the algorithm are the probabilities $\pi_r$ that a randomly selected node is assigned to class $r$, with 
$r = 1, 2, \ldots \mathcal{N}_c $,
and the set of  probabilities  $\theta_{rj}$ of having a  connection 
from a node belonging to class $r$ to a certain node $j$.
Assuming that the functions $\theta$ and $\pi$ are given, the probability ${\rm Pr}(A,g \vert \pi, \theta)$ of realizing the given graph under a node classification $g$, such that $g_i$ is the 
class that node $i$ has been assigned to,  can be written as 
\begin{equation}
{\rm Pr}(A,g \vert \pi, \theta) = \prod_{i}  \pi_{g_i} \left [ \prod_j 
\theta_{g_i, j}^{A_{ij}} \right ],  
\label{eqn:NL_LH}
\end{equation}
${\rm Pr}(A,g \vert \pi, \theta)$ is the likelihood to be maximized, but it turns out 
to be more convenient to consider its logarithm instead
\begin{equation}
\mathcal{L}(\pi,\theta) = \sum_i \left [ \ln \pi_{g_i} + \sum_j A_{ij} \ln \theta_{g_i, j} \right ]. 
\end{equation}
Treating the a priori unknown class assignment $g_i$ of the nodes 
as statistical "unknown data", one introduces next the auxiliary probabilities
$q_{ir} = {\rm Pr}(g_i \vert A, \pi, \theta)$ that a node $i$ is assigned to class $r$, and considers the averaged log-likelihood constructed as 
\begin{equation}
\bar{\mathcal{L}}(\pi,\theta) = \sum_{ir} q_{ir} \left [ \ln \pi_{r} + \sum_j A_{ij} \ln \theta_{rj} \right ].
\end{equation}
The maximization of $\bar{\mathcal{L}}$ must be performed taking into account the following normalization conditions for the probabilities $\pi$ and 
$\theta$
\begin{eqnarray}
\sum_{r=1}^{\mathcal{N}_c} \pi_r &=& 1 \label{eqn:pinorm}\\ 
\sum_{j=1}^{N} \theta_{rj} &=& 1 \label{eqn:thnorm}.  
\end{eqnarray}
The final results are
\begin{eqnarray}
\pi_r &=& \frac{1}{N} \sum_i q_{ir} \label{eqn:piNL}\\
\theta_{rj} &=& \frac{\sum_i A_{ij} q_{ir}}{\sum_i k_i q_{ir}},   
\label{eqn:thNL}
\end{eqnarray}
where $k_i$ is the total degree of node $i$. The still unknown probabilities $q_{ir}$ are then determined 
a posteriori by noting that 
\begin{equation}
q_{ir} = {\rm Pr} (g_i = r \vert A, \pi, \theta) = 
\frac{ {\rm Pr}(A,g_i=r \vert \pi, \theta)}{{\rm Pr}(A \vert \pi, \theta)},
\end{equation}
from which one obtains
\begin{equation}
q_{ir} = \frac{\pi_r \prod_j \theta_{rj}^{A_{ij}}}{\sum_s \pi_s \prod_j \theta_{rj}^{A_{ij}} }.
\label{eqn:qNL}
\end{equation}
Eqs.~(\ref{eqn:piNL}), (\ref{eqn:thNL}) and (\ref{eqn:qNL}) form a set of self consistent equations for 
$q_{ir}, \theta_{rj}$ and $\pi_r$ that any extremum of the expected log-likelihood 
must satisfy. 

Thus, given a graph $\mathcal{G}$, the 
EM algorithm consists of picking a number of classes $\mathcal{N}_c$ into which the nodes are 
to be classified and searching for solutions of 
Eqs.~(\ref{eqn:piNL}), (\ref{eqn:thNL}) and (\ref{eqn:qNL}).  These equations were derived by 
Newman and Leicht \cite{newman07}. They also showed that when applied to diverse type of networks the resulting $q_{ir}$ and $\theta_{rj}$ yield useful information 
about the  internal structure of the network. Note that only a minimal amount of a priori information is supplied: the number of classes $\mathcal{N}_c$ and the network.

\begin{figure}
\includegraphics[width=7cm]{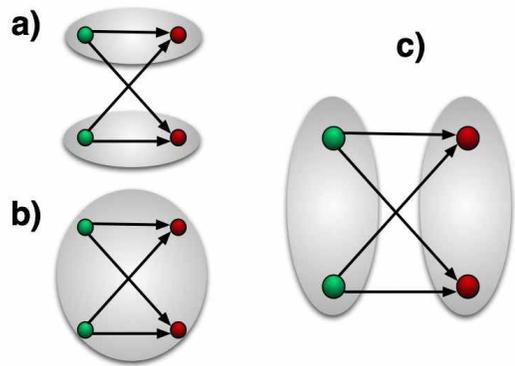}
\caption{A simple scenario in which the EM method for directed networks, as defined in \cite{newman07},  has problems to classify the nodes of the network in two classes. The configurations a) and b) are possible outputs of the original EM method since both satisfy the normalization condition of Eq.~(\ref{eqn:thnorm}).  The solution a) comes together with values for $q_{ir} = 1/2$ for all the nodes and classes, while the solution b), which has a lower likelihood, produces $q_{ir}\approx 0.99$  for all the nodes in one class and a very small probability for the other. The plot on the right, solution c), is the output offered by the generalization of EM with values of $q_{ir}$ virtually one and/or zero.}
\end{figure}

However, the EM method in the form presented so far does not yet serve our purposes for the following reason: as remarked before, content-based networks are usually represented as directed graphs. The probability $\theta_{rj}$ was defined as the probability that a node $j$ receives a directed connection from a node belonging to class $r$. Together with 
the normalization condition for $\theta_{rj}$, Eq.~(\ref{eqn:thnorm}), this implies
that the classification must be such that each class $r$ has at least one member 
with non-zero out-degree. This constraint forces the EM algorithm to classify a simple
bi-partite graph in the manner depicted in Figures 2a or 2b. From a content-based 
point of view, on the other hand, the classification that would be more natural 
is the one displayed Figure 2c which is forbidden by the condition of Eq.~(\ref{eqn:thnorm}). This difficulty is not resolved by re-defining $\theta_{rj}$ instead as the probability that a node $j$ 
makes a directed connection {\it to} a node belonging to class $r$, since now the classification must be such that each class $r$ has at least one member with non-zero in-degree.

We therefore have to generalize the EM approach in such a way that the node directionality 
does not restrict the possible classification of the nodes. This can 
be achieved by introducing the probabilities 
\begin{itemize}
\item $\ra{\theta}_{ri}$ of having a uni-directional link from a vertex of class $r$ to a node $i$, 
\item $\la{\theta}_{ri}$ of having a uni-directional link from node  $i$ to a node in class $r$, and  
\item $\ta{\theta}_{ri}$ of having a bidirectional link between $i$ and a node in class $r$.
\end{itemize}
 With these new definitions Eq. (\ref{eqn:NL_LH}) becomes
\begin{eqnarray}
&{\rm Pr}&(A,g|\pi,\la{\theta},\ra{\theta},\ta{\theta}) \\  
&=& \prod_i \left[ \pi_{g_i} \prod_j 
\la{\theta}_{g_i,j}^{A_{ji} \, (1-A_{ij}) } \,
\ra{\theta}_{g_i,j}^{A_{ij} \, (1-A_{ji}) } \,
\ta{\theta}_{g_i,j}^{A_{ij}\, A_{ji} } \right]. \nonumber
\label{prob2} 
\end{eqnarray}
The likelihood can now be written as
\begin{eqnarray}
\bar{\mathcal{L}}(\pi,\theta) &=& \sum_{ir} q_{ir} \left ( \ln{\pi_r} + \sum_j \left[ (A_{ji} \,
(1-A_{ij}) ) \ln{\la{\theta}_{r,j}}  \right. \right. \nonumber \\
&+& \left. \left. (A_{ij} \, (1-A_{ji})) \ln{\ra{\theta}_{r,j}} +  (A_{ij} \, A_{ji}) 
\ln{\ta{\theta}_{r,j}} ] \right] \right), \nonumber \\
\label{lbardirected}
\end{eqnarray}
which has to be maximized under the following constraint on the probabilities 
$\theta_{rj}$, 
\begin{equation}
\sum_i (\la{\theta}_{r,i}+ 
\ra{\theta}_{r,i} + \ta{\theta}_{r,i}) = 1,
\label{eqn:thdirnorm}
\end{equation}
implying that there is no isolated node. The probability $\pi_r$, that a randomly 
selected node belongs to class $r$, is again given by Eq.~(\ref{eqn:piNL}).  

Introducing the  Lagrange multipliers $\beta$ and $\lambda_r$, to incorporate the constraints,
Eqs.~(\ref{eqn:pinorm}) and (\ref{eqn:thdirnorm}), 
the expression to be extremized becomes 
\begin{eqnarray}
\tilde{\bar{\mathcal{L}}} &=&  \bar{\mathcal{L}} + \beta \left ( 1-\sum_r \pi_r \right ) \nonumber \\
&+& \sum_r \lambda_r \left( 1- \sum_i (\la{\theta}_{r,i}+ \ra{\theta}_{r,i} + \ta{\theta}_{r,i}) \right ).
\end{eqnarray}
As before, the extremal condition on $\tilde{\bar{\mathcal{L}}}$ with respect to 
$\pi$ gives us
\begin{equation}
\frac{\partial \tilde{\bar{\mathcal{L}}}}{\partial \pi_r} = 0 \Longleftrightarrow  \pi_r =
\frac{1}{N} \sum_i q_{ir}  \,\,\,\, \mbox{   and   } \,\,\,\, \beta = N ,
\end{equation} 
where $N$ is the total number of nodes. Differentiating $\tilde{\bar{\mathcal{L}}}$ with 
respect to the $\theta$ variables, we get \cite{smallnote}
\begin{eqnarray}
\frac{\partial \tilde{\bar{\mathcal{L}}}}{\partial \la{\theta}_{rj}} = 0 & \Leftrightarrow & 
\sum_i q_{ir} A_{ji} \, (1-A_{ij}) - \la{\theta}_{rj} \, \lambda_r = 0 \nonumber\\
\frac{\delta \tilde{\bar{\mathcal{L}}}}{\delta \ra{\theta}_{rj}} = 0 & \Leftrightarrow & 
\sum_i q_{ir} A_{ij} \, (1-A_{ji}) - \ra{\theta}_{rj} \, \lambda_r = 0 \nonumber \\
\frac{\delta \tilde{\bar{\mathcal{L}}}}{\delta \ta{\theta}_{rj}} = 0 & \Leftrightarrow & 
\sum_i q_{ir} A_{ij} \, A_{ji} - \ta{\theta}_{rj} \, \lambda_r = 0.  \nonumber \\
\label{thetaeqs}  
\end{eqnarray}
Putting together the three previous expressions and summing over the index of the
nodes $j$, we obtain the following result for the Lagrange multipliers 
\begin{equation}
\lambda_r = \sum_i q_{ir} \left( \bar{k}_i^i + \bar{k}_i^o - 
\bar{k}_i^b \right) ,
\end{equation} 
where $\bar{k}_i^i$, $\bar{k}_i^o$ and $\bar{k}_i^b$ are the in-degree,
out-degree and bi-directional degree of node $i$, respectively. Inserting
this relation into the previous set of equations, we extract the new extremal
conditions for the $\theta$'s
\begin{eqnarray}
\la{\theta}_{rj} = \frac{\sum_i q_{ir} A_{ji} \, (1-A_{ij})}{\sum_i q_{ir} ( \bar{k}_i^i + \bar{k}_i^o - 
\bar{k}_i^b )} \nonumber \\
\ra{\theta}_{rj} = \frac{\sum_i q_{ir} A_{ij} \, (1-A_{ji})}{\sum_i q_{ir} ( \bar{k}_i^i + \bar{k}_i^o - 
\bar{k}_i^b )} \label{eqn:thetarj} \\ 
\ta{\theta}_{rj} = \frac{\sum_i q_{ir} A_{ij} \, A_{ji}}{\sum_i q_{ir} ( \bar{k}_i^i + \bar{k}_i^o - 
\bar{k}_i^b )} \nonumber  .
\end{eqnarray}
These expressions have to be again supplemented with the self-consistent equation for
$q_{ir}$ which now reads 
\begin{equation}
q_{ir} = \frac{ \pi_r \prod_j 
\la{\theta}_{rj}^{A_{ji} \, (1-A_{ij}) } \,
\ra{\theta}_{rj}^{A_{ij} \, (1-A_{ji}) } \,
\ta{\theta}_{rj}^{A_{ij}\, A_{ji} }}{\sum_s \pi_s \prod_j 
\la{\theta}_{sj}^{A_{ji} \, (1-A_{ij}) } \,
\ra{\theta}_{sj}^{A_{ij} \, (1-A_{ji}) } \,
\ta{\theta}_{sj}^{A_{ij}\, A_{ji} }}.
\label{eqn:q_ir}
\end{equation}

Note that when we have only bi-directional links so that $A_{ij} = A_{ij}$, it follows 
from Eq. (\ref{eqn:thetarj}) that 
$\la{\theta}_{rj} = \ra{\theta}_{rj} = 0$. Thus we recover the original EM equations
under the identification $\theta_{rj} = \ta{\theta}_{rj}$. 

It is easily shown that the solutions of the EM equations, Eqs.~(\ref{eqn:piNL}), (\ref{eqn:thetarj}) and (\ref{eqn:q_ir}), are such that if two nodes $i$ and $j$ are 
structurally equivalent, {\it i.e.} $A_{ik} = A_{jk}$ as well as $A_{ki} = A_{kj}$, 
for all $k$ then they will be classified in the same manner:  $q_{ir} = q_{jr}$, 
and $\la{\theta}_{ri} = \la{\theta}_{rj}$, $\ra{\theta}_{ri} = \ra{\theta}_{rj}$ and $\ta{\theta}_{ri} = \ta{\theta}_{rj}$  for all $r$. This property of the solutions obtained from 
the EM methods renders it very well-suited for detecting any underlying 
contents-based structure.  

\section{The Inversion Method}

One important shortcoming of the EM method is that $\mathcal{N}_c$ has to be provided as an external parameter. The algorithm lacks a means to evaluate how good a classification is, and consequently one 
cannot decide which number of classes 
furnishes an optimal classification of the nodes of a graph. 
To overcome this problem, we propose to define a measure of the quality of a classification
as follows:
\begin{equation}
S_q = - \frac{1}{N} \sum_{i,r} q_{ir} \ln(q_{ir}),
\label{eqn:Sdef}
\end{equation}
where the sum runs over all the nodes $i$ and classes $r$. $S_q$ is the average entropy of the 
classification and as such measures the certainty with which the nodes are assigned to their respective 
classes. One can easily see that $0 \le S_q \le \ln \mathcal{N}_c$. For a sharp classification 
$S_q = 0$, while the worst-case scenario occurs when $q_{ir} = 1/\mathcal{N}_c$. 
We will later argue that $S_q$ is a useful statistic to infer $\mathcal{N}_c$. 

Once an optimal classification has been found, it is possible to determine the connectivity 
structure among the classes. Given an EM classification, we will define $\tilde{c}(r,s)$ as the probability 
that a node in class $r$ has a connection to one in class $s$. This probability can be estimated
as
\begin{equation}
\tilde{c}(r,s) = \frac{\sum_{ij} q_{ir} A_{ij}  q_{js}}{ \sum_i q_{ir} \, \sum_j q_{js}} \, \left( 1  + \frac{\delta_{rs}}{\sum_i q_{ir}-1} \right) , 
\label{eqn:ctilde}
\end{equation}
by noting that 
\begin{equation}
p(i|r) = \frac{q_{ir}}{\sum_j q_{jr} }
\label{eqn:pir}
\end{equation}
is the posterior probability that given that a node has been assigned to class $r$, the node is $i$. 
The second term on the right hand side of Eq.~(\ref{eqn:ctilde}) must be included as a 
correction for the absence of self-connections, since by convention, we assume 
that $A_{ii} = 0$ for all $i$. 

$\tilde{c}(r,s)$, as defined above, is the probability of regarding 
a connection between two nodes in the graph as being one between nodes of type $r$ 
and $s$. 
As we will show in the following section, if the underlying graph is a contents-based 
network, a successful application of the EM algorithm should result in sharp 
assignments of 
nodes into classes and $\tilde{c}(r,s)$ should thus be binary valued (and moreover be 
equal to the connectivity function $c(r,s)$).  
It is possible to also define a measure of how close the connectivity function 
resembles one that corresponds to a content-based network by considering the entropy for $\tilde{c}$, 
\begin{equation}
S_c = - \frac{2}{\mathcal{N}_c^2 \, \ln{2}} \; 
\sum_{rs} \tilde{c}(r,s) \ln \tilde{c}(r,s).
\label{eqn:sc}
\end{equation}
We have that $ 0 \le S_c \le 1$. The maximum of $S_c$ occurs when 
$\tilde{c}(r,s) = 1/2$, i.e. when none of the classes have any 
preferred connection pattern to any class.   

The generalization of the EM method, the entropies $S_q$, $S_c$ and the estimation of 
$\tilde{c}(r,s)$ are in general applicable to any kind of graph. However, for the 
purpose of this article we will focus only on their applications to content-based networks. 
We will address the  general case in a subsequent work \cite{MunganRamasco2007}, where we will also show that contents-based networks play a special role for the 
classifications of the EM method. 

\section{Analytical Results for Contents-Based Networks}

Assume that we are given a contents-based graph $\mathcal{G}$ that 
has been constructed from a set of nodes 
of unknown contents, and an unknown connectivity function $c(x,y)$. In this setting, we 
suppose that the optimal number of classes $\mathcal{N}_c$ has already been found and that 
it is equal to the number of contents $\mathcal{N}_x$.
We would like to know under which conditions the EM algorithm can infer 
the class membership of the nodes as well as the connectivity function. In other 
words, given the adjacency matrix $A_{ij}$, we are looking for a solution of 
the generalized EM equations, Eqs.~(\ref{eqn:thetarj}) and (\ref{eqn:q_ir}), with
\begin{equation}
q_{ir} = \delta_{r,x_i} \, \, \, \, \, \mbox{with} \, \, \, \, \,  x_i \in \mathcal{X}  ,
\label{eqn:leakfree}
\end{equation}
along with the unknown class-connectivity function $\tilde{c}(r,s)$ that ideally should 
coincide with the original $c(x,y)$. Note that the Ansatz Eq.~(\ref{eqn:leakfree}) 
implies that for such a solution  $S_q = 0$.

Substituting the above Ansatz into Eq.~(\ref{eqn:thetarj}), we find
\begin{eqnarray}
\la{\theta}_{rj} &=& \frac{ c(x_j,r) \, \left [ 1 - c(r,x_j) \right ] }{\bar{k}_r^i + 
\bar{k}_r^o - \bar{k}_r^b } \nonumber \\
\ra{\theta}_{rj} &=& \frac{c(r,x_j) \, \left [ 1 - c(x_j,r) \right ]}{\bar{k}_r^i + 
\bar{k}_r^o - \bar{k}_r^b } \label{eqn:theqs}  \\
\ta{\theta}_{rj} &=& \frac{c(r,x_j) c(x_j,r) }{\bar{k}_r^i + \bar{k}_r^o - \bar{k}_r^b}, 
\nonumber
\end{eqnarray}
 where $\bar{k}_r^i$, $\bar{k}_r^o$ and $\bar{k}_r^b$ are the average in-degree, 
out-degree and bi-directional degree of nodes belonging to class $r$, 
\begin{eqnarray}
\sum_i \delta_{x_i,r} \left ( \bar{k}_i^i + \bar{k}_i^o - \bar{k}_i^b \right ) &=& 
n_r \, \left ( \bar{k}_r^i + \bar{k}_r^o - \bar{k}_r^b \right ) \nonumber \\
&\equiv& n_r \, \bar{k}_r, 
\end{eqnarray}
so that $\bar{k}_r$ is the total degree of each of the $n_r$ nodes belonging to class $r$. 
Note that in Eq.~(\ref{eqn:theqs}), 
the node index $j$ enters only through its content $x_j$, so that $\theta_{rj}$ 
is the same 
for all the nodes that have the same content as $j$. The same turns out to be true for the 
$q_{ir}$. We thus have  $q_{ir} = q_{tr}$ for all nodes $i$ such that $x_i = t$, and 
from Eq.~(\ref{eqn:q_ir}) we obtain
\begin{eqnarray}
q_{tr} = \frac{\gamma_{t} \pi_r}{\bar{k}_r^{ \bar{k}_t}} \; &\prod_s& \left \{
\left [ c(r,s) \, (1-c(s,r))  \right ]^{c(t,s) \, (1-c(s,t))}   \right. \nonumber \\ 
&\times&  \left [ c(s,r) \, (1-c(r,s))  \right ]^{c(s,t) \, (1-c(t,s)) } \, \nonumber \\
&\times& \left. \left [ c(r,s) \, c(s,r) \right ]^{c(t,s)\, c(s,t) } \right \},
\label{eqn:cbcondition}
\end{eqnarray}
where $\gamma_{t}$ is the normalization constant for $q_{tr}$.

We now have to consider the conditions on $c(r,s), c(s,r), c(t,s)$, and $c(s,t)$ such that 
given the classes $r$ and $t$, the terms in the product on the 
right hand side of Eq. (\ref{eqn:cbcondition}) are non-zero for all $s$, when 
$r=t$, and zero for at least one $s$ when $r\ne t$. This is a statement about the kind of 
connections that the nodes of type $r$ and $t$ make to or receive from nodes of all 
possible classes $s$.
An inspection of the $c^c$ type terms in the product shows that their contribution 
to $q_{tr}$ is non-zero if and only if the following two conditions are satisfied for all $s$: 
\begin{itemize}
\item If there is a 
connection between $t$ and $s$, there must be also a connection between $r$ and $s$  of 
the same kind, namely either in, out, or bi-directional. 
\item Whenever there is no connection between $t$ and $s$, 
there can be any kind of connection  between $r$ and $s$, as well as none at all.
\end{itemize} 
The satisfaction of both conditions can be regarded as constituting a cover type of 
relation between $r$ and $t$, {\it i.e.} nodes belonging to class $r$
connect in the same way with all the classes that nodes belonging to class $t$ connect, 
{\it but} they have also some extra connections. We will denote this relation by $r \succ 
t$ and say that $r$ covers $t$. From its definition it is clear that the cover relation 
is transitive, $r \succ t, t \succ s \Longrightarrow r \succ s$. When 
$r \succ t$, we also define   $\mathcal{E}(r; t)$ as the set of extra 
classes that $r$ connects 
to (or receives connections from) relative to those of $t$. 

With the above definition, it can be readily seen that when $r \succ t$  
\begin{equation}
\bar{k}_r = \bar{k}_t + \sum_{v \in \mathcal{E}(r; t)} \, n_v,
\end{equation}
where the index $v$ runs over the extra classes to which $r$ is connected. 
This implies that
\begin{equation}
\bar{k}_r^{ \bar{k}_t} = \bar{k}_t^{ \bar{k}_t} \, \left ( 1 + \frac{\sum_{v \in 
\mathcal{E}(r; t)} \, n_v}{\bar{k}_t} \right )^{\bar{k}_t} .
\end{equation}
Thus we find that 
\begin{eqnarray}
q_{tr} = \left \{ \begin{array}{ll} 
\frac{\gamma_t \pi_t}{\bar{k}_t^{ \bar{k}_t}}  & r = t, \\ 
\frac{\gamma_t \pi_r}{\bar{k}_t^{ \bar{k}_t}} \left ( 1 + \frac{\sum_{v \in \mathcal{E}(r; 
t)} \, n_v}{\bar{k}_t} \right )^{-\bar{k}_t} & r \succ t, \\ 
0 & \mbox{o/w.} \end{array} \right.
\label{eqn:qleak}
\end{eqnarray}
(with $\mathcal{E}(t; t) \equiv \emptyset $). Thus, when $r \succ t$ and for large  
$\bar{k}_t$, $q_{tr}$ deviates from our Ansatz, Eq.~(\ref{eqn:leakfree}), by an 
exponentially small amount.

Treating the deviations 
caused by the presence of cover relations among the classes, as a small perturbation 
to our Ansatz, Eq.~(\ref{eqn:leakfree}), we can obtain the leading order expression 
for $q_{tr}$ as 
\begin{eqnarray}
q_{tr} = \left \{ \begin{array}{ll} 
1 - \sum_{r \succ t} \frac{\pi_r}{\pi_t} 
\left ( 1 + \frac{\sum_{v \in \mathcal{E}(r; t)} \, n_v}{\bar{k}_t} \right )^{-\bar{k}_t}
 & r = t ,\\ 
\frac{\pi_r}{\pi_t} \left ( 1 + \frac{\sum_{v \in \mathcal{E}(r; t)} \, n_v}{\bar{k}_t} 
\right )^{-\bar{k}_t} & r \succ t, \\ 
0 & \mbox{o/w,} \end{array} \right.
\label{eqn:qContBased}
\end{eqnarray}
where $\gamma_t$ has been determined from the normalization 
\begin{equation}
\sum_r q_{tr} = 1.   
\end{equation}

To the same order, we find also that
\begin{eqnarray}
\pi_r = \frac{n_r}{N} &-& \sum_{t \succ r} \frac{n_t}{N} \left ( 1 + \frac{\sum_{v \in 
\mathcal{E}(t; r)} \, n_v}{\bar{k}_r} \right )^{-\bar{k}_r} \nonumber \\
&+& \sum_{t \prec r} \frac{n_r}{N} \left ( 1 + \frac{\sum_{v \in \mathcal{E}(r; t)} \, 
n_v}{\bar{k}_t} \right )^{-\bar{k}_t}.
\label{eqn:piContBased}
\end{eqnarray}
Equations (\ref{eqn:qContBased}) and (\ref{eqn:piContBased}) are the analytical solution 
of the EM equations for a content-based network with connectivity function $c(r,s)$.

We see that whenever a class $r \succ t$, 
there is a non-zero probability for a node $t$ to be also classified as belonging to class 
$r$. We will refer to this as a leakage in the class assignment. 
However as can be seen from Eq.~(\ref{eqn:qContBased}), the leakage 
probabilities vanish exponentially with the size of the classes. 
A detailed account of the solution structure for 
contents-based networks as well as more general types of networks will be given 
elsewhere \cite{MunganRamasco2007}. 

\begin{figure}
\includegraphics[width=6cm]{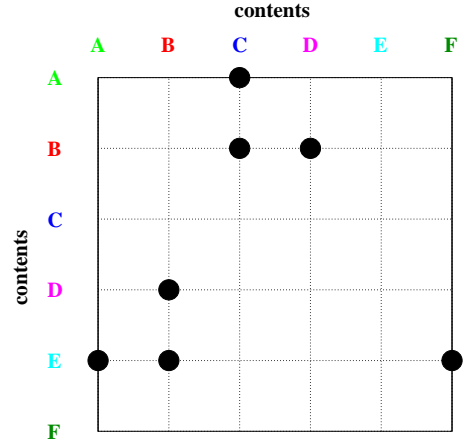}
\caption{Connectivity function $c(x,y)$ for the theoretical example of Section V-A. The 
number of contents 
is six: $A, B, C, D, E$ and $F$. The points represent the ones in the connectivity matrix, 
the values not marked are zero.}
\end{figure}

When the contents-based network is cover-free, the generalized EM equations have a 
leak-free solution
and thus the entropy of the class assignments $S_q$ vanishes. On the other hand, in the 
presence of cover relations, the EM method will produce 
assignments with some nodes in multiple classes, i.e. leaks.  We have already found above
the leading order behavior for the leakage. It is not too hard to show that, in that case, 
$S_q$ is 
given by
\begin{equation}
S_q =  \sum_{t \mbox{\footnotesize{ has a cover}}}\sum_{r \succ t} n_r \alpha(r;t) \, \left( 1 + \frac{\alpha(r;t)}{\bar{k}_t} \right)^{-\bar{k}_t},  
\label{eqn:Sqcb}
\end{equation}
where $\alpha(r;t) \equiv \sum_{v \in \mathcal{E}(r;t)} n_v$ is the number of nodes  
to which nodes in class $r$ are connected in addition to those that nodes in class $t$ connect. In many practical situations, the number of contents is 
fixed. This implies that if the probability of being in class $r$ is given by $p_r$, the actual number of nodes in the $r$ class 
will grow on average as $n_r = N \, p_r$ with the system size. Thefore, the factor $ \alpha(r;t)$ of Eq.~(\ref{eqn:Sqcb}) can also be written as
\begin{equation}
\alpha(r;t) = N \, \tilde{\alpha}(r;t) , 
\end{equation}
where $\tilde{ \alpha}(r;t)$ is a constant depending on the connectivity function that generated the network. Under these assumptions, the entropy $S_q$ will decrease exponentially with the network size, meaning that even for moderately sized networks 
the leakages will be in general too small to cause significant 
misclassification. 

As shown in Section IV, the solution of the EM 
equations provides us with an estimate for the class connectivity, $\tilde{c}(r,s)$, given by Eq.~(\ref{eqn:ctilde}).
For contents-based networks  
in the absence of any cover relation among classes, we have, 
{\it cf.} Eq.~(\ref{eqn:pir}), $p(i\vert r) = 
\delta_{x_i,r}/n_r$. and from 
Eq.~(\ref{eqn:ctilde}) we immediately find that $\tilde{c}(r,s) = c(r,s)$ with $S_c = 0$.  
In the presence of cover relations among the classes, there will be corrections that 
vanish exponentially with the number of nodes in the relevant classes.
These results demonstrate that the EM algorithm is 
capable of inferring the hidden class connectivity 
function that generated the network.

\subsection{An Example}

\begin{figure}[b!]
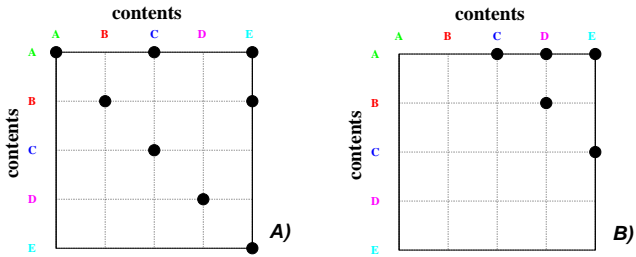

\includegraphics[width=3.8cm]{fig4a.eps}
\qquad
\includegraphics[width=3.8cm]{fig4b.eps}
\caption{Connectivity functions $c(x,y)$ for the two examples of content-based networks analyzed in the simulation sections. The number of contents considered is five, $A,B,C,D$ and $E$. The contents of the connectivity function $\mathcal{A})$ display no cover relation, while in the second example, $\mathcal{B})$, $A \succ B$. The networks are generated assuming equal probability for the five contents at the assignation of a content to each node.}
\end{figure}

In order to further illustrate the theoretical results above, we turn next to an example. 
Consider a network generated from six kinds of contents to be denoted by $A, B, C, D, E$ 
and $F$, and 
with the connectivity function as shown in Figure 3. 
The following cover relations are present: 
$B \succ A \succ F$; that is, $B \succ A$, $B \succ F$, and $A \succ F$. In fact, we have 
chosen this particular example to elucidate the effect of having nested covers and to show 
that the cover relation is transitive. 
For each of the cover relations, the sets of connections to additional classes are  
$\mathcal{E}(B; A) = \{D \}$,  $\mathcal{E}(B; F) = \{D, C\}$ and  $\mathcal{E}(A; F) = 
\{C\}$. When inserted into Eq. (\ref{eqn:qContBased}), these relations yield
\begin{eqnarray}
q_{AA} &=& 1 - \frac{n_B}{n_A} \left ( 1 + \frac{n_D}{n_E + n_C} \right )^{-n_E - n_C}, \nonumber \\
q_{AB} &=& \frac{n_B}{n_A} \left( 1 + \frac{n_D}{n_E + n_C} \right )^{-n_E - n_C},  \\
q_{FF} &=& 1 - \frac{n_A}{n_F} \left ( 1 + \frac{n_C}{n_E} \right )^{-n_E}  
- \frac{n_B}{n_F} \left ( 1 + \frac{n_C + n_D}{n_E} \right )^{-n_E}, \nonumber \\
q_{FA} &=& \frac{n_A}{n_F} \left ( 1 + \frac{n_C}{n_E} \right )^{-n_E}, \nonumber \\
q_{FB} &=& \frac{n_B}{n_F} \left ( 1 + \frac{n_C + n_D}{n_E} \right )^{-n_E}, \nonumber 
\label{eqn:leakage_exp}
\end{eqnarray}
with $q_{BB} = q_{CC} = q_{DD} = q_{EE} = 1$ and all the other values of $q_{rt} = 0$. 
These results are in agreement with what one would expect intuitively. For example, since 
$B \succ A$ and $B \succ F$, there is a non-zero probability of mistaking nodes of type 
$A$ or $F$ by nodes of $B$, i.e. $q_{AB}$, $q_{FB}$, and $q_{FA}$ are all non-zero. However 
this probability vanishes exponentially with the number of nodes in the classes $E$ and $C$. 
In the large network size limit, the leakage on $q_{ir}$, and how far $S_q$ deviates 
from zero, are determined by the pair of classes $(r,t)$ such 
that $r$ is the "tightest" cover of $t$, these are the pairs $r \succ t$ for which 
$\alpha(r;t)$ is smallest,  $A \succ F$ and $B \succ A$, in our example.

\begin{figure}[b!]
\includegraphics[width=3.9cm]{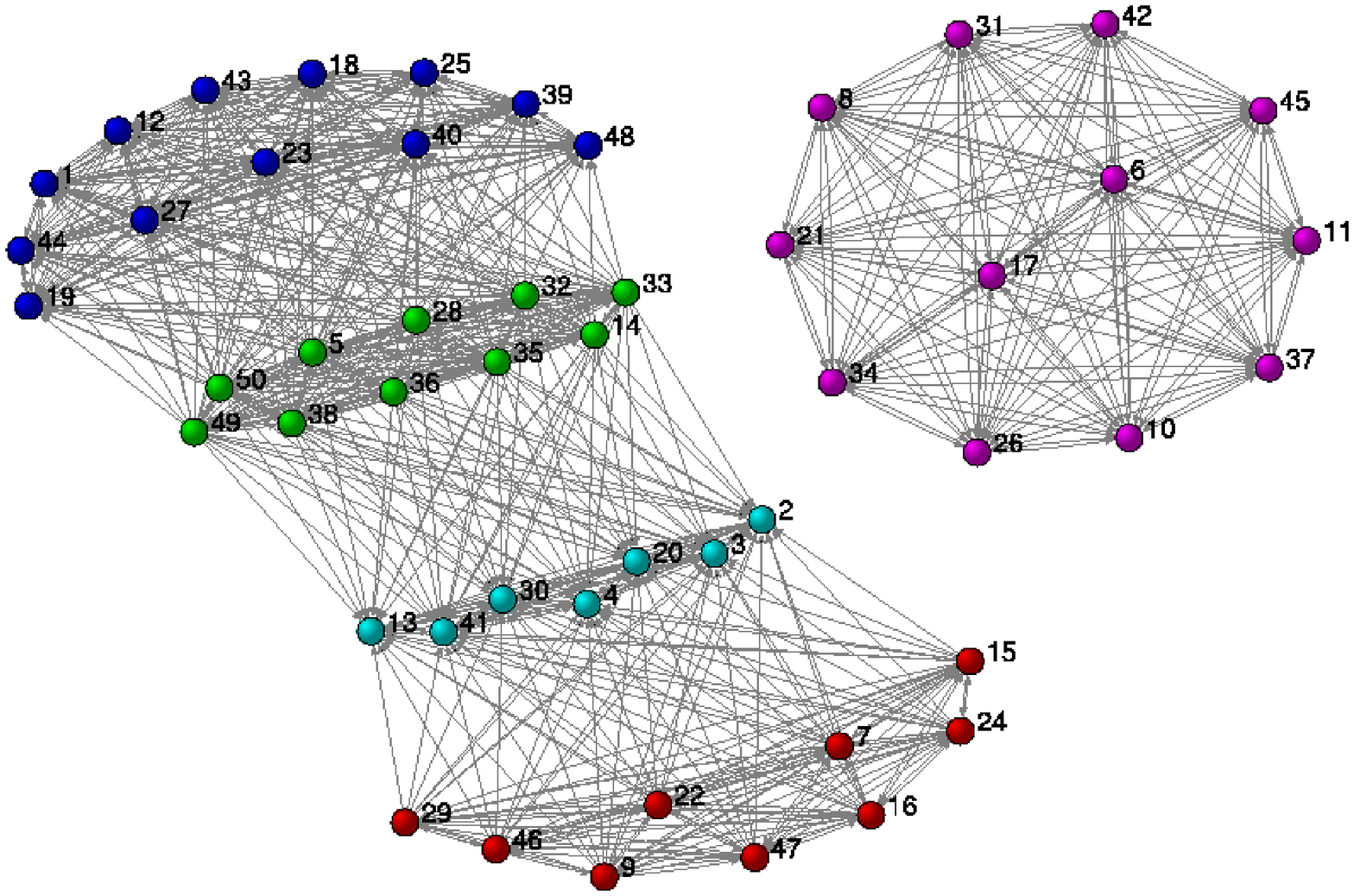}
\qquad
\includegraphics[width=3.9cm]{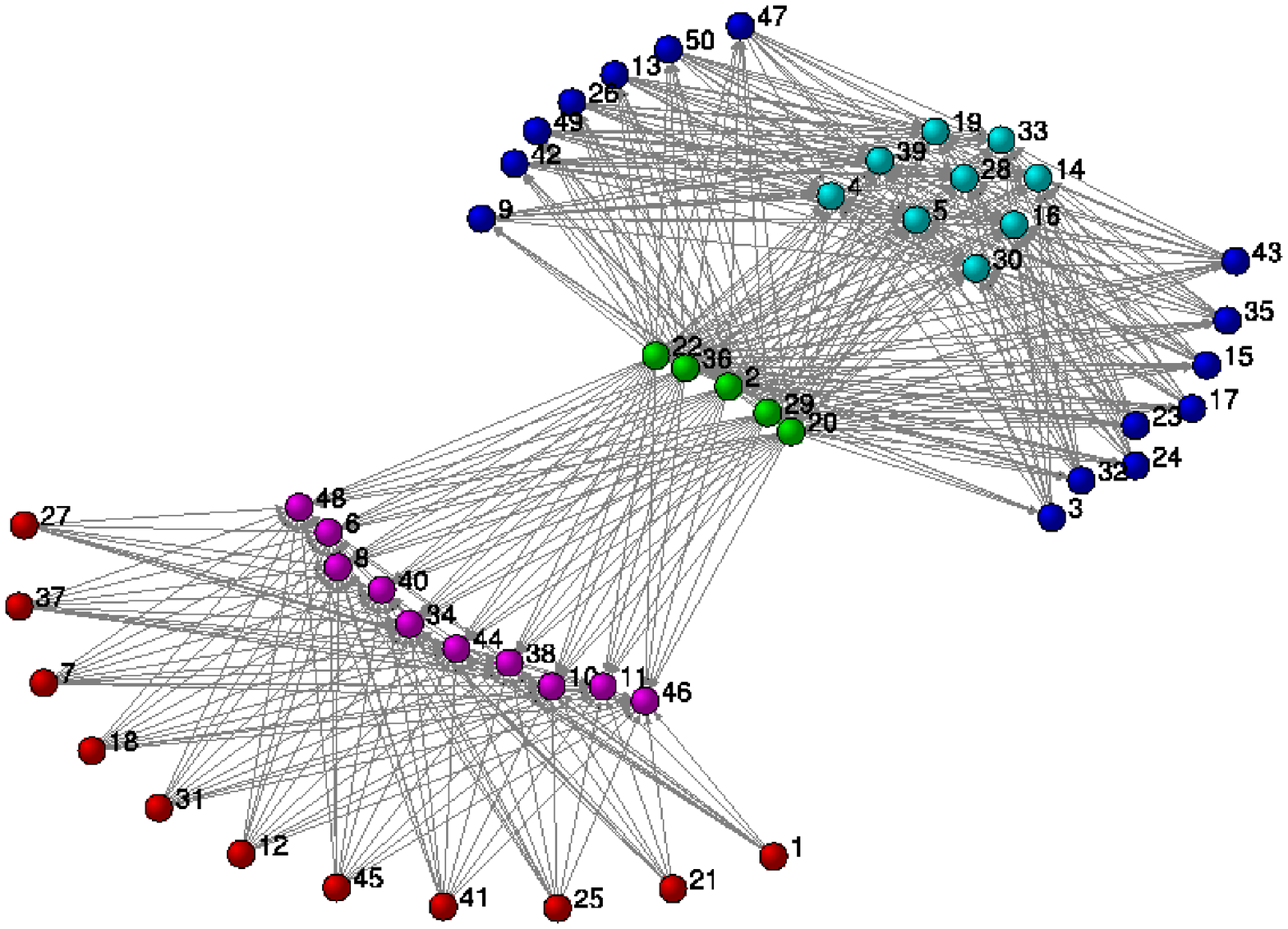}
\qquad
\includegraphics[width=3.9cm]{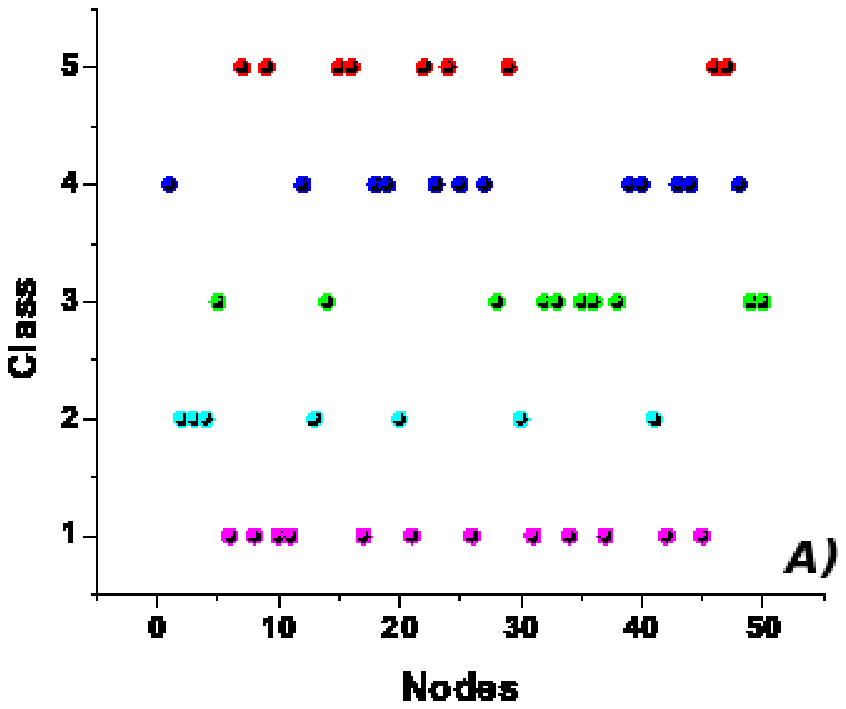}
\qquad
\includegraphics[width=3.9cm]{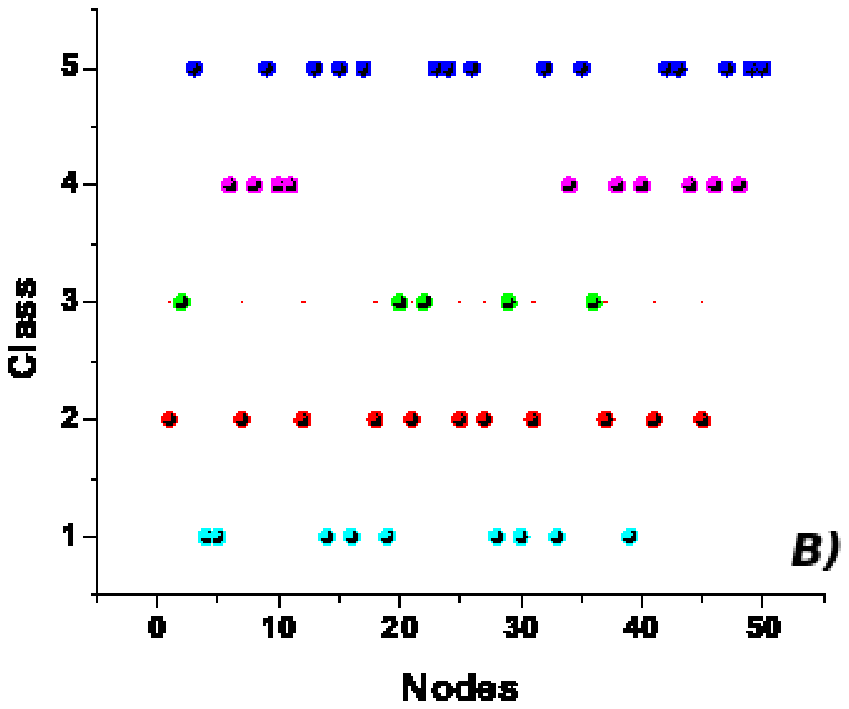}
\caption{An example of classification, the original network is on the top and
on the bottom the probability $q_{ir}$ is represented graphically. The color of 
the symbols correspond to the contents of the nodes (green $A$, red $B$, magenta
$C$, blue $D$ and cyan $E$). On the bottom, the spheres radius
is proportional to the probability $q_{ir}$. On the left, the network is generated using the connectivity function $c_{\mathcal{A}}$ of Figure 4 with no cover relation among the 
classes, while on the right we have used $c_{\mathcal{B}}$, which incorporates a single cover relation between $A$ and $B$ such that $A \succ B$.}
\end{figure}

\section{Simulation Results, EM applied to Content-Based Networks}

In the following, we study numerically the ideas introduced in the previous 
sections. The generalized version of EM will be applied to directed
content-based networks generated randomly from the connectivity functions shown in Figure 4. The nodes of these networks have a content assigned that is selected at random out of $\mathcal{N}_x = 5$ five possible contents, denoted by $A, B, C, D$ and $E$. Since the presence of coverage relations can change the quality of an EM classification, we have considered two connectivity functions $c(x,y)$ (see Fig. 4); one without class coverage, $c_\mathcal{A}$, and another, $c_\mathcal{B}$, with a single cover relation 
between contents $A$ and $B$, such that $A \succ B$. In order to improve our 
numerical 
estimation of the classification with maximum likelihood, we implemented a 
simulated-annealing type of procedure for the optimization of 
$\overline{\ml}$.

\begin{figure}
\includegraphics[width=8.cm]{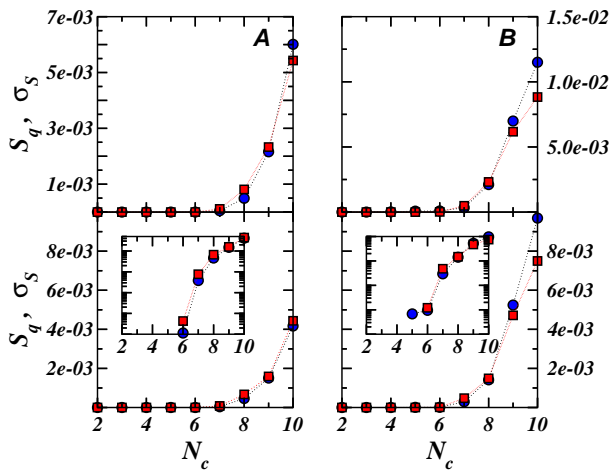}
\caption{In the low panels $S_q$ (circles) and its fluctuations $\sigma_S$ (squares) as a function of $\mn_c$ for the networks shown in Figure 5. In order to facilitate visualization, the insets show the same curves in a semi-logarithmic plot. The top panels display 
the same quantities, $S_q$ and $\sigma_S$, but ensemble-averaged over different realizations of the content-based networks generated with the connectivity function of Figure 4.}
\end{figure}

In the previous section, we have shown that our generalized EM method is able 
to infer the 
underlying content-based structure that generated the network. These 
calculations were carried out assuming that the number of contents 
$\mathcal{N}_x$ coincides with the number of classes $\mathcal{N}_c$. Let us 
therefore start by setting  $\mathcal{N}_c= \mathcal{N}_x = 5$. In Figure 5, 
we show graphically the classifications obtained from the generalized EM 
method as applied to an ensemble of networks of size $N = 50$ generated with 
the connectivity functions of Figure 4. The color coding is based on the 
contents of the nodes and will be such that it  matches in all the subsequent 
figures of the paper ($A$ green, $B$ red, $C$ blue, $D$ magenta and $E$ cyan). 
The size of the spheres in the bottom plots are proportional to the 
probabilities $q_{ir}$. For these examples the classification is rather good 
even in the case when a cover relation is present, as can be readily seen 
from the bottom diagrams where no major color is misplaced. In other words, 
there 
are not misclassifications, although for the $\mathcal{B}$ case a small amount
of leakage can be noticed. 

To try to quantify the quality of these results, we can, as a first measure, count the number of network realizations in our ensemble for which at least two nodes with different contents have been assigned to the same class, with the understanding that a node $i$ is assigned to a class 
$r$ whenever $q_{ir} > 1/2$. This is a strict criterion, since it may well be that we are considering as {\em erroneous} a classification with only a single node misclassified. The result can also slightly depend on the method applied to optimize the likelihood. Still, this definition is a way to play on safe ground and avoid to complicate too much the detection of mistakes in the classification. Let us call this then the error 
rate of the classification $\epsilon$. For each of the two connectivity functions of Fig. 4, we  have studied over
$2000$ realizations of networks of size $N = 50$. In none of them the generalized EM method misclassified a single node. 
This result is in agreement with our earlier observation that the EM method classifies 
structurally equivalent nodes in the same way. 

\begin{figure}
\includegraphics[width=8.cm]{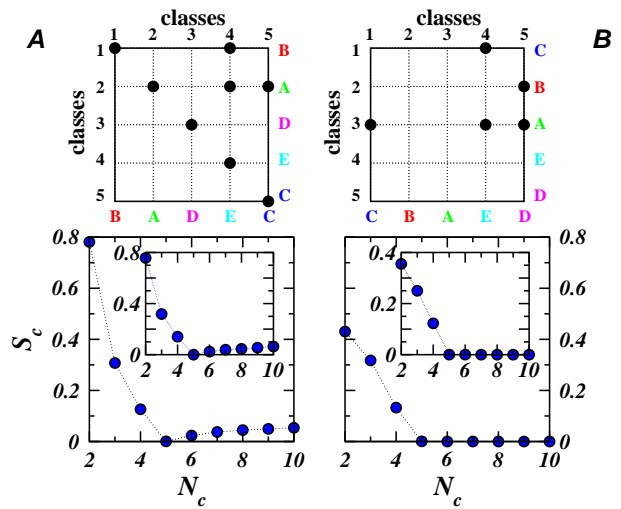}
\caption{On the top, the connectivity function $\tilde{c}(r,s)$ obtained from the EM classification of the networks displayed in Figure 5. The radii of the circles is proportional to the value of $\tilde{c}(r,s)$.  On the bottom, we are showing how $S_c$ goes with $\mn_c$ for the same networks as well as, in the inset, an average over different content-based realizations generated with the connectivity functions of Fig. 4.}
\end{figure}

The next question is then: how can the optimal $\mathcal{N}_c$ be determined? If
the networks studied are content-based, there are several possible answers to
this. Here we will outline two of them and will discuss at the end of this section a third one in the context of inferring the class connectivity function. In Section
IV, we have introduced a measure $S_q$ for the quality of an EM
classification of the network. We have also shown that when $\mathcal{N}_x =
\mathcal{N}_c$, $S_q$ is either zero or exponentially small for large content-based networks.
Therefore, a signal on $S_q$ can be expected for $\mathcal{N}_c =
\mathcal{N}_x$, 
if the EM algorithm is faced with the challenge  of classifying a content-based
network with a series of values $\mathcal{N}_c$. 
This effect happens because the normalization conditions of Eqs. (\ref{eqn:pinorm}) and
(\ref{eqn:thdirnorm}) impose that no class can be left totally unassigned, 
 $\pi_{r} > 0$ for all $r$. The more redundant classes the
method has to assign nodes to, the higher $S_q$ becomes. 
In other words, we are providing the EM algorithm with a larger 
degree of freedom than required to properly classify the nodes.
The extra freedom leads to structural leakage. The evolution of $S_q$ with
$\mathcal{N}_c$ is displayed in Figure 6 for the two networks of Fig. 5. These
are, of course, particular examples but some general features can be deduced.
First, the value of $S_q$ is rather small or even zero for $\mathcal{N}_c < \mathcal{N}_x$. 
This may be a generic property of content-based networks. As noted before, the 
structural equivalence 
of nodes with the same content prevents the EM algorithm from putting such 
nodes into different 
classes. This means that once
the contents are classified by classes the leakage comes from cover relations
between classes and can become very small for big networks.
 On 
the other 
hand, when $\mathcal{N}_c > \mathcal{N}_x$, the availability of excess classes that cannot 
be left totally unassigned causes $S_q$ to be non-zero and to increase steadily with 
$\mathcal{N}_c$. The boundary between these two types of behaviors is precisely the unknown 
$\mathcal{N}_c = \mathcal{N}_x$.

Another peculiarity of the EM method applied to contents-based networks is that when  $\mathcal{N}_c < \mathcal{N}_x$, the landscape of the likelihood seems
to have a very clear and unique maximum. The solution at the point of maximum 
$\overline{\mathcal{L}}(\pi,\theta)$ has also a well determined value of $S_q$. However, if
$\mathcal{N}_c \ge \mathcal{N}_x$, the landscape of the likelihood becomes 
 rough, with a large number of local maxima. The search for the global maximum under
such conditions is therefore much harder. And even, in the cases where it can be numerically found, say when $\mathcal{N}_c = \mathcal{N}_x$, it is formed by a set of
degenerate extrema with the same value of $\overline{\mathcal{L}}$ but very
different values of $S_q$. Indeed, the values of the entropy shown in Fig. 6 for
$\mathcal{N}_c \ge \mathcal{N}_x$ are averages over the best likelihood
solutions found in different realizations of the optimization methods along with 
their standard deviations $\sigma_S$. The dispersion $\sigma_S$, 
of $S_q$ around its average, can be used in practice as another estimator for the
optimal number of classes (see Fig. 6).

\begin{figure*}
\centering
\includegraphics[width=5.cm]{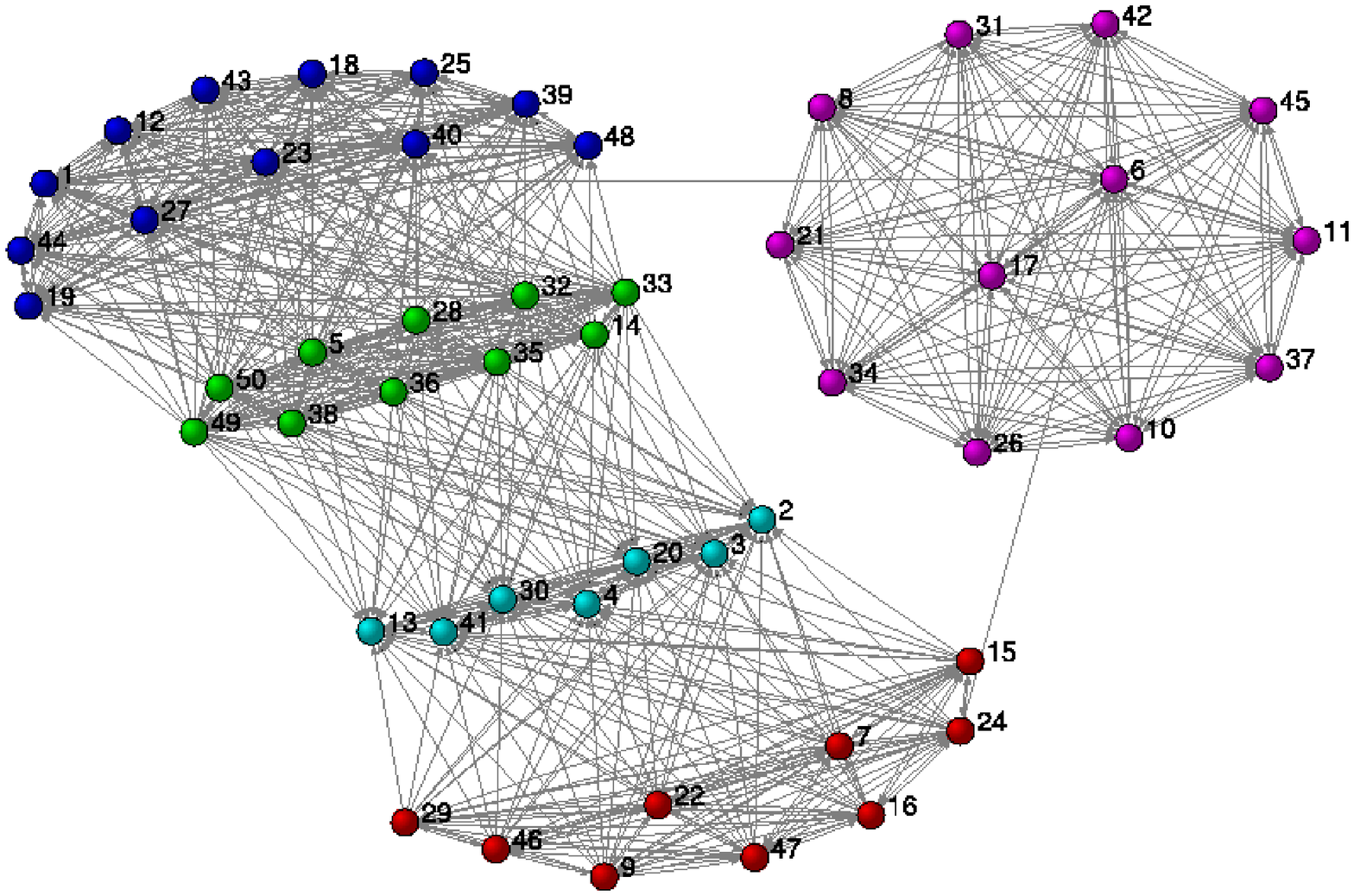}
\qquad
\includegraphics[width=5.cm]{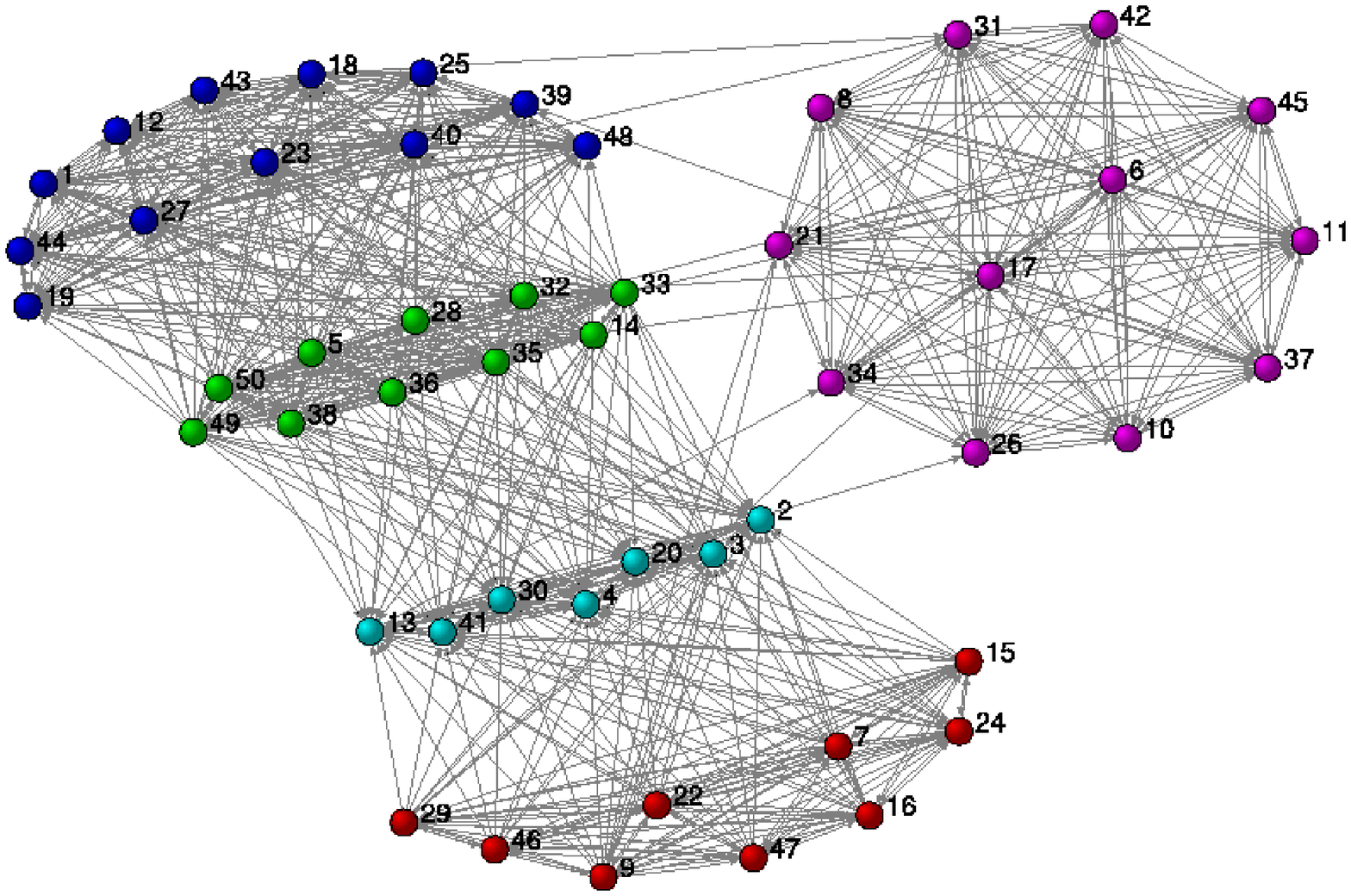}
\qquad
\includegraphics[width=5.cm]{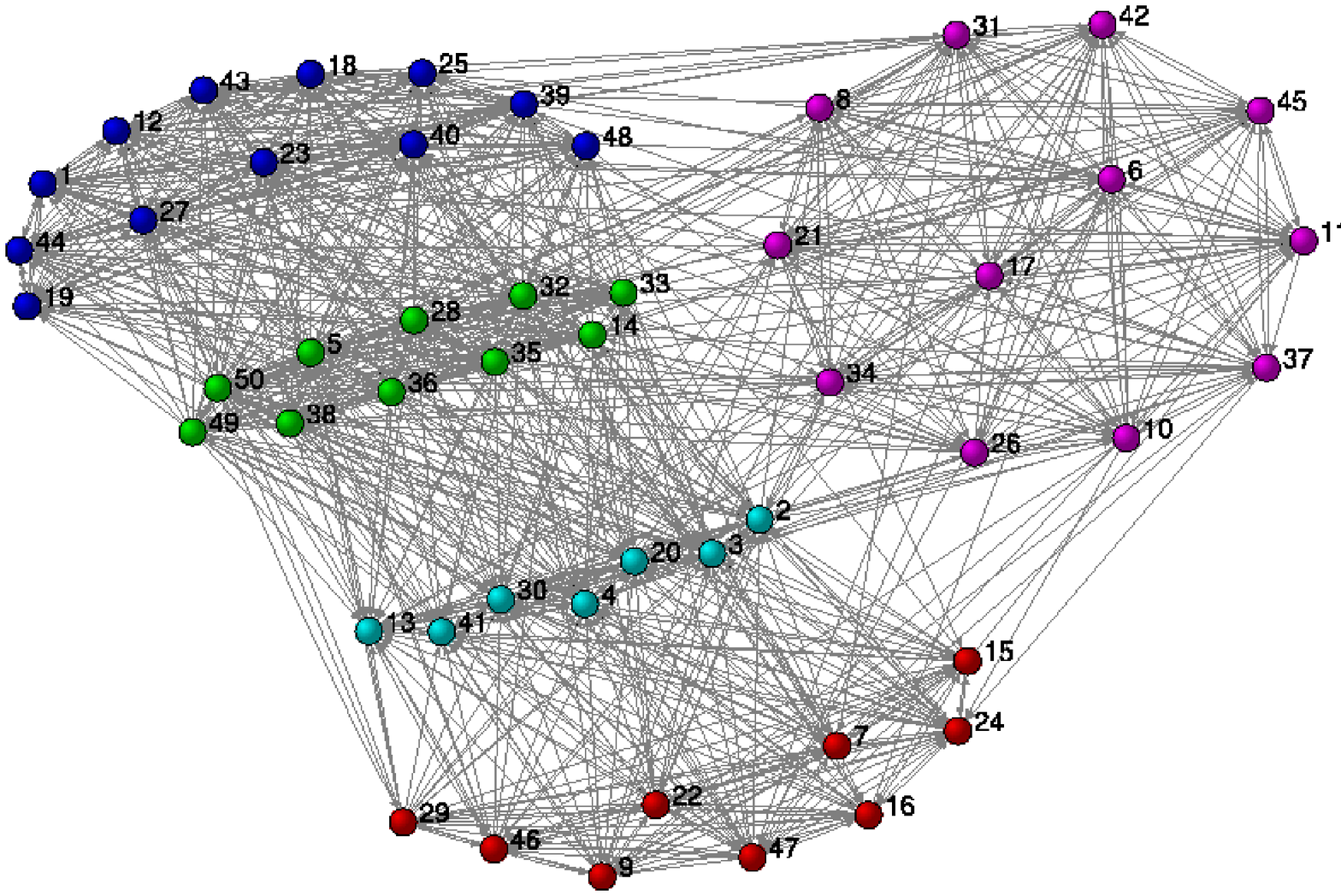}
\qquad
\includegraphics[width=5.cm]{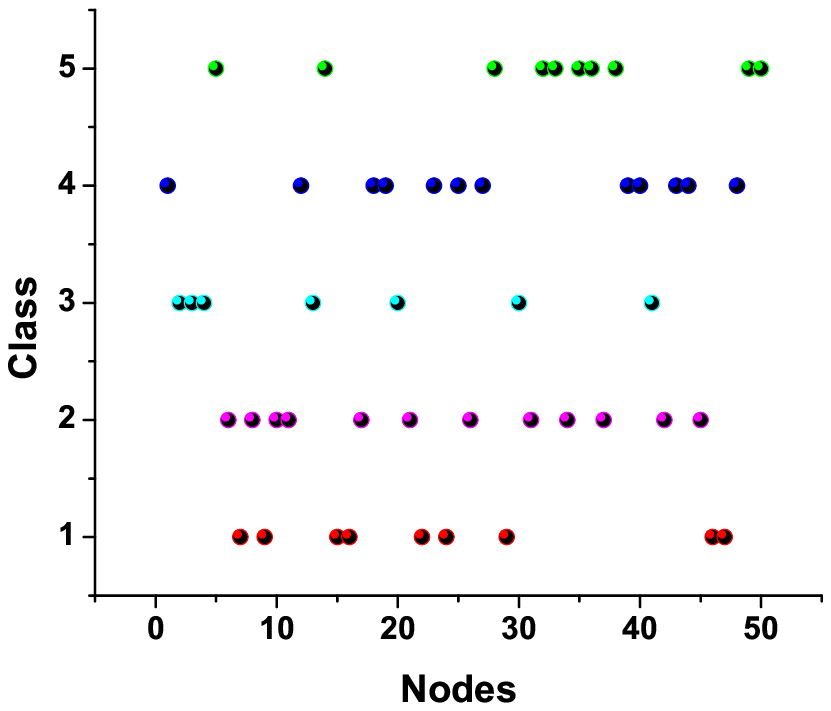}
\qquad
\includegraphics[width=5.cm]{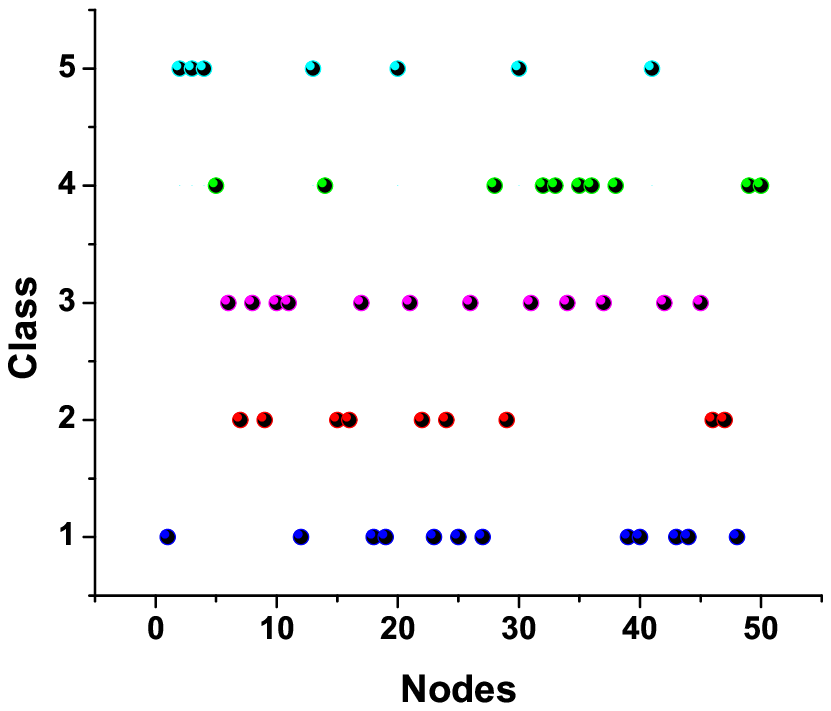}
\qquad
\includegraphics[width=5.cm]{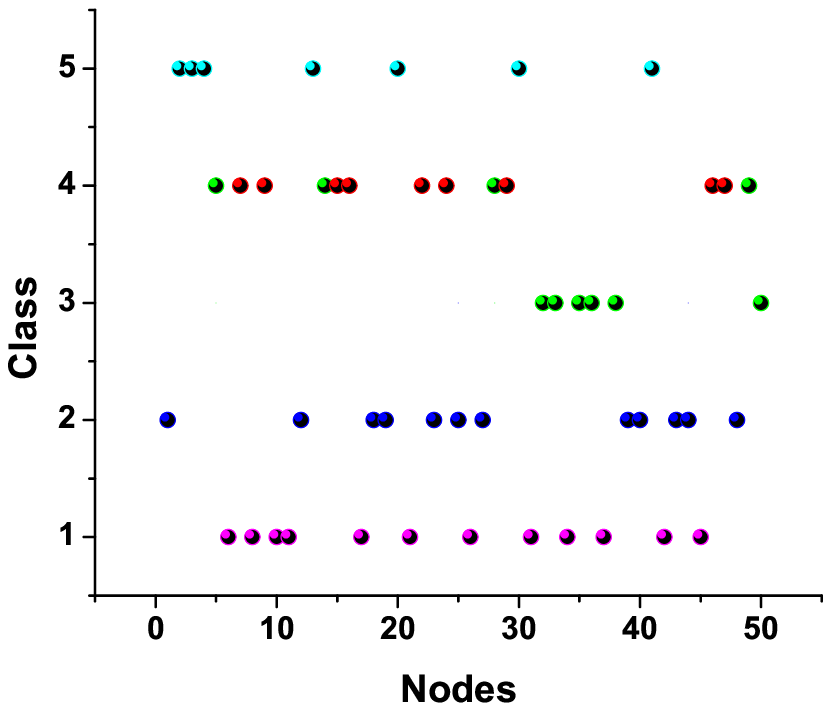}
\caption{Same network as in Section $\mathcal{A}$ of Fig. 5 but with increasing error probability $P_\alpha = P_\mu$. The
values of $P_\alpha$ are from left to right $0.001, 0.01$ and $0.1$. The
plots on the lower panel are a graphic representation of the probability of classifying 
node $i$ in class $r$, $q_{ir}$, as before the radius of the spheres are proportional to $q_{ir}$ and the colors correspond to the actual content of the nodes (green $A$, red $B$, magenta
$C$, blue $D$ and cyan $E$).}
\end{figure*}

Once $\mathcal{N}_x$ is known, it is possible to recover $c(r,s)$ as explained
in Section IV. In the top panels of Figure 7, the recovered $\tilde{c}(r,s)$ is displayed for the
content based networks of Figure 5. After the classes of $\tilde{c}(r,s)$ have
been properly reordered, it is impossible to distinguish the top panels of Fig.
7 from the connectivity functions given in Fig. 4. Also, in the lower panels of
Figure 7, we have included the evolution of the entropy $S_c$ as a function of
$\mathcal{N}_c$. $S_c$ also shows a clear change of behavior at $\mathcal{N}_c =
\mathcal{N}_x$, suggesting that the best content-based partition of the network
happens when the number of classes equals the number of contents. Consequently, $S_c$, apart from being an estimator of how much a network deviates from a purely content-based graph, 
is also a useful quantity for deciding when $\mathcal{N}_c$ is optimal.

\begin{figure}[b!]
\includegraphics[width=8.cm]{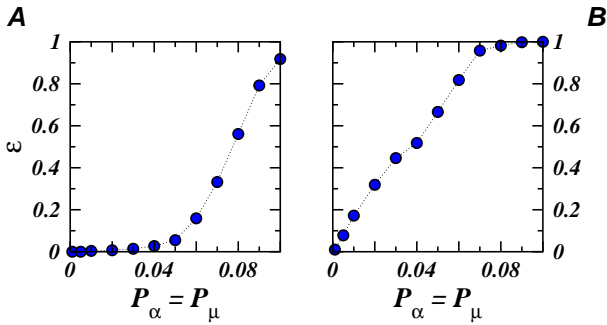}
\caption{The error rate $\epsilon$ as a function of the probabilities $P_\alpha = P_\mu$ for content-based networks generated with the connectivity functions of Figure 4 and with $\mn_c = \mn_x = 5$.}
\end{figure}

\section{EM and Noisy connections in Content-Based Networks}

It is unlikely that in real-world networks the generating processes is error-free. 
Even if the underlaying structure is expected to be a content-based network, 
errors in the connecting pattern could naturally arise. We try to mimic  
the unexpected connections as well as the absence of expected connections, by introducing 
the corresponding error probabilities to the process of network generation from 
its contents. As before, each node $i$ has a
content $x_i$ assigned at random from the set of possible contents (in the 
case of our example networks the same five possibilities: $A, B,C, D$ and $E$). 
Once the contents are established, 
the structure of the content-based network should be determined completely by the 
connectivity function $c(x_i,x_j)$: If $c(x_i,x_j)= 1$, there ought to be a link 
from node $i$ to $j$, and none if $c(x_i,x_j) = 0$. As a way of gradually loosing the content-based structure of the connections, we introduce now the probabilities $P_\mu$, and $P_\alpha$,  of not having a link, when $c(x_i,x_j) = 1$ and having a link although $c(x_i,x_j) = 0$, respectively. 
The networks constructed in this way can be regarded as hidden variable graphs \cite{caldarelli,soderberg,marian} for which the probability of connection between any nodes 
$i$ and $j$ is expressed as 
\begin{equation}
r(x_i,x_j) = c(x_i,x_j) \, (1 - P_\mu)  + [1 - c(x_i,x_j)] \, P_\alpha .
\label{eqn:hidden}
\end{equation}
In other words, where in the absence of noise the probability of having a connection was 
one, it now is $1-P_\mu$, and likewise, where it was zero, it now is $P_\alpha$. The extreme 
limit of this model occurs when  $P_\mu = P_\alpha = 1/2$, so that the probability of connecting to a node of other class is maximally random and independent of the connectivity function. We are more interested here in the limit when both $P_\alpha$ and $P_\mu$ are much smaller than $1/2$, and the resulting graphs can be seen as a slight modification of a content-based network. For the sake of simplicity, all of the results shown below are for $P_\alpha = P_\mu$.

\begin{figure}
\includegraphics[width=8.cm]{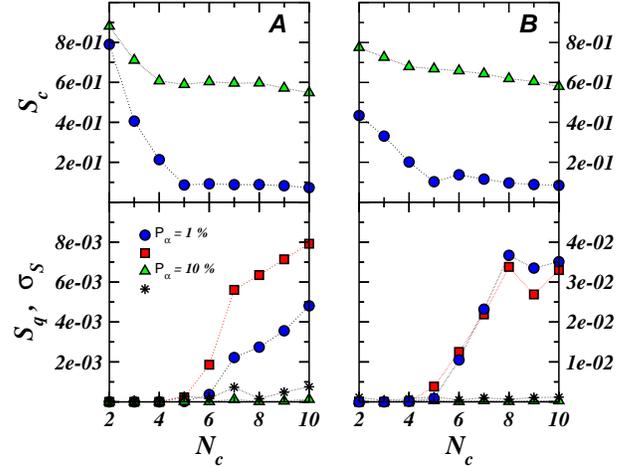}
\caption{The average entropies over different realizations for content-based networks generated with the connectivity functions of Figure 4. In the top panels, $S_c$ is represented as a function of the number of classes $\mn_c$ for two different levels of disorder: the circles are $P_\alpha = P_\mu = 1 \%$, while the triangles for $P_\alpha = P_\mu = 10 \%$. On the bottom panels, $S_q$ and $\sigma_S$ versus $\mn_c$ for 
the disorder probabilities $P_\alpha = P_\mu = 1 \%$, circles ($S_q$) and squares ($\sigma_S$), and $P_\alpha = P_\mu = 10 \%$, triangles ($S_q$) and stars ($\sigma_S$).}
\end{figure}

Let us begin by looking at how the networks change with increasing assignment
error. In the top panels of Figure 8, we display a series of networks generated with the connectivity function $c_\mathcal{A}$ for different values $P_\mu = P_\alpha$. 
It is readily seen that the connection patterns associated with the different kinds of 
content becomes more and more diffuse. On the bottom panels of the same figure, we 
show the corresponding class assignment probabilities $q_{ir}$. While these are just 
examples, there are some features that are worth pointing out. The problems in the 
classification seem to appear somewhere between $P_\alpha = P_\mu = 1 \%$ and $P_\alpha = 
P_\mu = 10 \%$. Even at $10 \%$ of error the number of nodes misclassified in these networks 
is not very high. A closer inspection of the solution found shows that actually only two of 
the node contents-classes are mingled up, while all the remaining node classes are perfectly 
assigned. With the aim of quantifying these 
observations, the behavior of $\epsilon$ is plotted in Figure 9 versus the disorder probability. 
This plot is, of course, susceptible to slight changes depending on the method used to 
search for the maximum likelihood and depends on how many realizations of the content-based graphs were considered (in this case $1000$). Nevertheless, in our simulations the threshold for a sharp classification of
all the nodes of the network is around $P_\alpha = P_\mu \approx 5 \%$
for graphs without coverage, connectivity function $c_\mathcal{A}$, and much 
lower, around $P_\alpha = P_\mu \approx 2 \%$, for those with a cover 
relation,  $c_\mathcal{B}$. The exact value will depend on the particular 
connectivity function, apart from the optimization method, but these values 
give us already an idea about the order of magnitude of the threshold beyond 
which the content-based structure cannot be recovered anymore.

\begin{figure}[b!]
\includegraphics[width=8.cm]{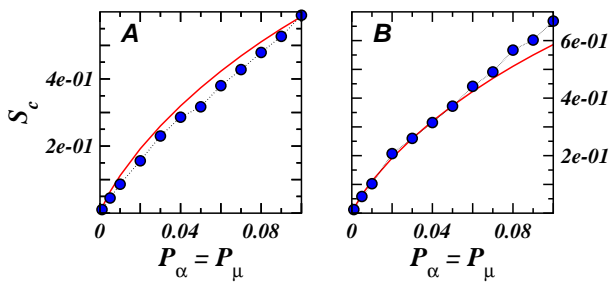}
\caption{The average entropy $S_c$ as a function of the disorder probability $P_\alpha = P_\mu$ for 
content-based networks generated with the connectivity functions $c_\mathcal{A}$ and $c_\mathcal{B}$ depicted in Figure 4. The red curves correspond to the value of $S_c^*$.}
\end{figure}

A next aspect to consider is how the entropies $S_q$ and $S_c$ are affected by the 
intensity of the disorder, and whether they are still valid estimators to determine the 
optimal number of classes. To answer this question, we fix the probabilities $P_\mu = 
P_\alpha$ at $1 \%$, which seems to be a value where one might plausibly expect to obtain 
good classifications for both type of networks. In Figure 10, we display $S_q$, $\sigma_S$ 
and $S_c$, as function of the number of classes $\mn_c$ with the results averaged over 
different content-based realizations. Indeed, at this level of disorder the entropies can still be used to  
estimate $\mn_x$. The noise in the connections introduces a small constant background for $S_c$, 
which we will denote by $S_c^*$, and which can be determined in both examples from the 
behavior at high values of $\mn_c$. We can estimate the value of $S_c^*$ by noting that  
when $\mn_c = \mn_x$ any non-zero entropy should essentially be due to the background 
from the random connections. Substituting the expression for $r(x_i,x_j)$, 
Eq. (\ref{eqn:hidden}), into the definition of $S_c$, Eq.~(\ref{eqn:sc}), should 
therefore give us an estimate for $S_c^*$,
\begin{equation}
S_c^* \approx - \frac{2}{\mn_c^2 \, \ln{2}} \sum_{x,y} r(x,y) \ln{r(x,y)}
\label{eqn:sc_star}
\end{equation}
For $P_\alpha = P_\mu = 1 \%$, this yields $S_c^*\sim 0.112$, close to the value observed in the Figure 10 for $\mn_c \ge 5$. 
To check how well our estimate for $S_c^*$ agrees with the values obtained from 
simulations, we plot in Figure 11 $S_c$ vs. the disorder probability at 
$\mn_c = 5$. When the disorder becomes very strong, on the other hand, it 
might not be possible to find an optimal 
$\mn_c$. Moreover, the presence of very different connection patterns for 
nodes with the same content renders the existence of such optimal number 
dubious. Therefore, apart from the obvious 
classification $\mn_c = N$, there may not be any other sharp classification. 
The effects of high disorder can be seen in Figure 10, where the entropies 
$S_c$ and $S_q$  are represented as a function of $\mn_c$ for 
$P_\alpha = P_\mu = 10 \%$. The results depend on the connectivity function, 
$c_\mathcal{A}$ seems a little more robust to the disorder as was confirmed by 
Fig. 9, but the signal in $S_q$ or $\sigma_S$ is clearly lost or has moved 
to higher values of $\mn_c$. Also $S_c$ has lost its capacity to predict 
$\mn_x$ and smoothly falls for higher and higher values of $\mn_c$. It is 
worthy also noting that in spite of the lack of a method to find $\mn_x$, 
if $\mn_c = 5$ the EM method retrieves the appropriate hidden variable theory 
connectivity function $r(x,y)$ as can be inferred from the good fit produced 
by Eq. (\ref{eqn:sc_star}) to $S_c$ shown in Figure 11.  

The numerical findings of this section show that the classifications of the 
EM method are robust to the introduction of noise in the connection patterns up to a certain point.
The certainty of the classification will suffer, the stronger the disorder becomes. In fact 
this is one of the major merits of the EM method: it is able to extract the underlying 
content-based structure even in the presence of a certain level of noisy connections.

\section{Conclusions}

In summary, we have shown how the EM method for the classification of nodes of a network can be 
applied to 
content-based networks in order to  extract the underlying content-based  
structure even in the presence of a certain level of disorder in the connections. The application of 
the EM method to content-based networks is a natural concept 
that follows from the observation that the EM method classifies structurally equivalent 
nodes in an identical manner. In this sense, the EM method can be related to the Block Modeling techniques proposed in Social Sciences. Content based networks, on the other hand,  
are of great relevance, since they can be 
regarded as idealized paradigms of networks with 
communities or multipartite structures, including mixtures of both. Since in many 
realistic graphs the vertices carry additional attributes 
which might influence or even determine their connections to other vertices, being able 
to extract any content-based pattern can 
provide information about how the networks emerged. 

Our approach in this paper has been to start out with pure content-based graphs,
and to show analytically as well as numerically that the EM method can infer the 
content-based connectivity pattern. We have shown also that the existence of cover-relations 
between contents leads to non-zero probabilities of mistaking nodes belonging to different classes. However, these probabilities vanish exponentially with the increasing number of 
nodes, {\it i.e.}, the more discriminating information provided to the method. By regarding more realistic 
networks as perturbations of content-based networks under the addition or removal 
of connections, we then asked under which circumstances the EM method is still able to perform satisfactorily. There is a certain level of disorder beyond which the inference of the content-based structure, specially the number of contents, becomes rather hard if not impossible. 

In order to estimate the quality of the classification and how far the structure of the network is from a content-based structure, we have introduced two entropies, $S_q$ and $S_c$, which actually can be useful for the classification of any kind of graphs, including real-world networks. We have also shown that these entropies are applicable to deduce the optimal number of classes needed by the EM method to obtain a sharp classification of the nodes of the network.

{\it Acknowledgments---}  The authors would like to thank Alessandro Vespignani, Santo Fortunato, Filippo Radicchi, and in general the members of the Cx-Nets collaboration for useful discussions and comments. Funding from the Progetto Lagrange of the CRT Foundation, the Research Fund of 
Bo\u gazi\c ci University, as well as the Nahide and Mustafa Saydan Foundation was received. In addition, MM would like to acknowledge the kind hospitality of the ISI Foundation.


\begin{thebibliography}{99}
\bibitem{smallnote} The equations written in this form take care of the 
case when some of the $\theta$ are zero, as can be readily checked by comparing 
their solution, Eq. (\ref{eqn:thetarj}), with Eqs. (\ref{lbardirected}) and (\ref{thetaeqs}).
\end{thebibliography}
\end{document}